\documentclass[jgrga]{agutex2015}
\usepackage{lineno}
\usepackage{textcomp}
\usepackage[varg]{txfonts}
\usepackage [pdftex]{graphicx}
\usepackage{float}
\authorrunninghead{GUO ET AL.}
\titlerunninghead{VARIATION OF THE RADIATION ON MARS}

\begin{document}

\title{Dependence of the Martian radiation environment on atmospheric depth: modelling and measurement}

\authors{ Jingnan Guo, \altaffilmark{1}
 Tony C. Slaba, \altaffilmark{2}
 Cary Zeitlin,\altaffilmark{3}
 Robert~F. Wimmer-Schweingruber,\altaffilmark{1}
 Francis~F Badavi,\altaffilmark{4}
 Eckart B\"ohm,\altaffilmark{1}
 Stephan B\"ottcher,\altaffilmark{1}
 David E. Brinza,\altaffilmark{5}
 Bent Ehresmann,\altaffilmark{6}
 Donald~M. Hassler,\altaffilmark{6}
 Daniel Matthi\"a, \altaffilmark{7}
 Scot Rafkin\altaffilmark{6}}
\altaffiltext{1}{Institute of Experimental and Applied Physics, Christian-Albrechts-University, Kiel, Germany. {guo@physik.uni-kiel.de}}
\altaffiltext{2}{NASA Langley Research Center, Hampton, VA, USA}
\altaffiltext{3}{Lockheed Martin IS and GS, Oakland, CA, USA}
\altaffiltext{4}{Old Dominion University Research Foundation, Norfolk, VA, USA}
\altaffiltext{5}{Jet Propulsion Laboratory, California Institute of Technology, Pasadena, CA, USA}
\altaffiltext{6}{Southwest Research Institute, Space Science and Engineering Division, Boulder,CO, USA}
\altaffiltext{7}{Aerospace Medicine, Deutsches Zentrum f\"ur Luft- und Raumfahrt, K\"oln, Germany}

\keypoints{\item Anticorrelation between Martian surface radiation and atmospheric column depth 
\item The influence of this correlation by solar activities}

\setpagewiselinenumbers

\begin{abstract}
The energetic particle environment on the Martian surface is influenced by solar and heliospheric modulation and changes in the local atmospheric pressure (or column depth). 
The Radiation Assessment Detector (RAD) on board the Mars Science Laboratory (MSL) rover Curiosity on the surface of Mars has been measuring this effect for over four Earth years (about two Martian years). 
The anti-correlation between the recorded surface Galactic Cosmic Ray (GCR) induced dose rates and pressure changes has been investigated by \citet{rafkin2014} and the long-term solar modulation has also been empirically analyzed and modeled by \citet{guo2015modeling}. 
This paper employs the newly updated HZETRN2015 code to model the Martian atmospheric shielding effect on the accumulated dose rates and the change of this effect under different solar modulation and atmospheric conditions. 
The modeled results are compared with the most up-to-date (from 14 August 2012 until 29 June 2016) observations of the RAD instrument on the surface of Mars. 
Both model and measurements agree reasonably well and show the atmospheric shielding effect under weak solar modulation conditions and the decline of this effect as solar modulation becomes stronger.
This result is important for better risk estimations of future human explorations to Mars under different heliospheric and Martian atmospheric conditions.
\end{abstract}

\begin{article}

\section{Introduction and motivation}\label{sec_intro}
In order to plan future human missions to Mars 
the assessment of the radiation environment on and near the surface of Mars is necessary and fundamental for the safety of astronauts \citep{cucinotta2011updates}. 
However, contributions to the radiation environment on the Martian surface are very complex \citep[e.g.,][]{saganti2002}: energetic particles, such as galactic cosmic rays (GCRs) and solar energetic particles (SEPs), entering the Martian atmosphere may create secondary particles via spallation and fragmentation processes\footnote{Particles with low charges and greater mean free paths may also pass through the $\sim$ 22 g/cm$^2$ of atmosphere without any {nuclear} interactions with the ambient atomic nuclei}, which may further interact while propagating through and finally result in {very complex spectra} when reaching the surface of Mars. 
The downward particle flux reaching the Martian surface may also interact in the regolith and, amongst other outcomes, produce backscattered {particles} which can be detected on the surface or in orbit \citep[e.g.,][]{boynton2004}. 

In case of radiation on the Martian surface, the planet itself serves as a good shielding against interplanetary energetic particles that would come from the bottom half of the full solid angle. 
For the top half, the Martian atmosphere shielding is dependent on the zenith angle: the column depth in the vertical direction is much smaller than toward the horizon. The first determination of this zenith angle dependence showed that the radiation field coming from within zenith angles of up to $\sim$ 15 degrees at Gale crater is most isotropic with slight increase of shielding towards the larger zenith angles \citep{wimmer2015}.  
The Martian atmosphere exhibits a strong thermal tide excited by direct solar heating on the day side and strong infrared cooling on the night side. Heating causes an inflation of the atmosphere with a simultaneous drop in surface column density \citep{rafkin2014}. 
This daily variation of the column density results in a daily variation of the shielding depth. Moreover, the seasonal CO$_2$ condensation cycle leads to a seasonal pressure variation which also affects the atmospheric column depth in the long term\citep[e.g.,][]{guo2015modeling}.

There are two types of primary particles reaching the top of the atmosphere of Mars: GCRs and SEPs. 
SEPs, containing mainly protons and electrons, are sporadic and their intensity may differ greatly from case to case. 
The Martian atmosphere serves as a natural low-energy cutoff for incoming particles (about 200 MeV for protons) and only SEP events with a strong high-energy component can be seen on the surface.
GCRs are the main contribution to the surface Martian radiation and are modulated by the heliospheric magnetic field which evolves dynamically as solar activity varies in time, with a well-known 11-year cycle.
During the past 4 years of measurement, RAD has seen variations of the surface dose rates driven by both pressure changes and solar modulation concurrently. 
The anti-correlation between the recorded surface GCR dose rates and pressure (which can be converted to accumulated atmospheric column density) changes has been investigated by \citet{rafkin2014} using hourly-binned data to analyze the diurnal perturbations. 
The long-term solar modulation of the surface dose rates has also been empirically analyzed and modeled by \citet{guo2015modeling}. 

In the current paper, we discuss the atmospheric shielding effect of the GCR dose rates using the newly updated HZETRN2015 code and compare the results with observations of MSL/RAD on the Martian surface from the past four years.

\section{HZETRN2015 code and the modeling results}
The newly updated HZETRN2015 \citep{slaba2016solar, wilson2016} code was used in this work to model the radiation environment on the Martian surface over a broad range of input conditions. The code allows various levels of transport approximation to be considered from the highly efficient one-dimensional straight-ahead solution to more complex three-dimensional (3D) transport for neutrons and light ions \citep{wilson2014advances}.  
In this work, the bi-directional transport model was utilized for neutrons and light ions. This corresponds to the 3D transport calculation with N = 2 in the notation of \citet{wilson2014advances}. 
{I.e., incident GCR ions are modeled with an isotropic distribution and transported in straight lines; secondary ions are transported along their initial velocity vectors; secondary neutrons can be transported bidirectionally parallel or anti-parallel to their initial velocity vectors. }
Previous verification studies on the lunar surface \citep{slaba2011variation} and validation studies on the Mars surface \citep{matthia2016martian} have shown this transport model to be reasonably accurate compared to Monte Carlo simulations. Details of the nuclear physics models used in HZETRN2015 can be found in \citet{wilson1991, wilson2014advances, wilson2016}. 

\subsection{Modeling setup}

\begin{figure}
\noindent\includegraphics[width=0.80\columnwidth]{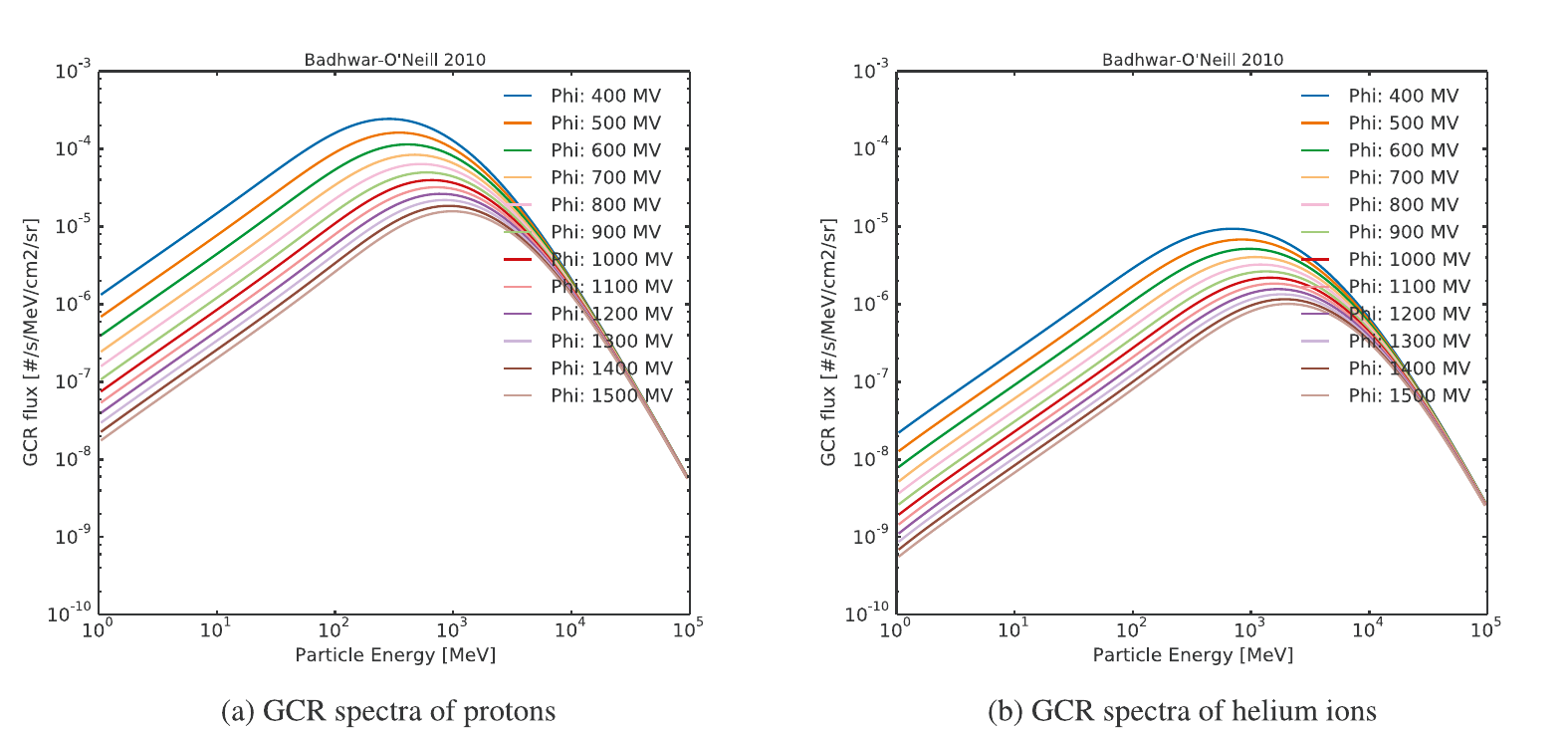}
\caption{GCR proton (top) and helium ion (bottom) spectra in the interplanetary space generated by the BON10 model (see text for description) under different values of solar modulation potential $\Phi$.
}\label{fig:modulation}
\end{figure}

Of particular interest in this work was to utilize the efficiency of HZETRN2015 to model surface radiation environment over a broad range of solar modulation potential $\Phi$ and atmospheric column density $\sigma$ values.
For solar modulation effects, the Badhwar-O'Neill 2010 \citep[BON10,][]{badhwar2010} model was used to generate GCR spectra of ion species ranging from proton to Ni (boundary condition for HZETRN2015) for $\Phi$ ranging from 400 MV to 1500 MV, corresponding to solar minimum and solar maximum, respectively. 
An example of the proton and helium ion (the most abundant $^4$He isotope is considered in this work) particle spectra are shown in Figure \ref{fig:modulation} over the range of $\Phi$ values. 
For atmospheric column density effects, the Mars Climate Database (MCD) version 4.3 \citep{millour2008mars} was used to define the vertical density profile near the Curiosity landing site at Gale crater. 
A range of solar longitudes and local times were evaluated in MCD to obtain seven different vertical thicknesses from 18.9 g/cm$^2$ to 25.7 g/cm$^2$, corresponding to the seasonal variation of the atmosphere measured by MSL \citep[e.g.,][]{guo2015modeling}.
The atmosphere was represented by a composition of 95\% CO$_2$, 2.7\% N$_2$, 1.6\% Ar as well as trace amounts of O$_2$ and CO \citep{deangelis2004}. The regolith was represented by a composition of about 47\% oxygen and 24\% silicon \citep{mckenna2012characterization} and set to be 1.75 meters thick (or 300 g/cm$^2$).

Using each of the GCR boundary conditions and atmospheric profiles as input into HZETRN2015, particle fluence, dose, and dose equivalent are computed on the surface using a ray-by-ray transport methodology {following \citet{slaba2013radiation} where Figure 1 sketches the cartoon geometry of the atmosphere model. 
In this approach, the atmosphere is represented as a spherical shell surrounding a solid sphere representing Mars. 
The thickness of the atmosphere shell is defined by the vertical density profile and then evaluated along a large number of rays covering the upper 2$\pi$ solid angle using geometric relationships \citep{simonsen1990radiation}.} 
Bi-directional transport is executed along each of the rays, including the atmospheric thickness and 300 g/cm$^2$ of martian regolith, and the total radiation field is obtained by integrating the individual ray-results. 
Results within a specified solid angle field of view may also be obtained by simply integrating over the ray-results falling inside the cone of interest. 
This approach has been shown to provide reasonably accurate spectral results for various particles compared to Monte Carlo simulation and MSL/RAD measurements \citep{matthia2016martian}, at least in the context of known uncertainties associated with nuclear physics models used in radiation transport codes \citep{norbury2012review}. 

Along with the surface results for each of the solar modulation and atmospheric input conditions, additional calculations were performed at various elevations above the surface. 
For these additional calculations, the evaluation point was viewed in the model as being positioned on the surface {with the lower layers of the atmosphere neglected but the solid regolith kept intact}. 
This enabled an even broader range of atmospheric shielding conditions to be evaluated, as will be shown – all of which utilize the same basic geometric setup and assumptions. 
{Given the small column density of the omitted atmosphere in comparison with that of the typical depth in solid material through which such particles would have been transported, such geometry approximation should have minimal impact on the updated neutron spectrum.} 
For all calculations, the resulting particle spectra were used to calculate total dose rate in both silicon and water materials. Proton and helium ion {(of both primary and secondary ones at the target point)} induced dose rates are also recorded separately. The dose equivalent rate was computed using the linear energy transfer (LET) dependent quality factor from ICRP 60 \citep{icrp60}.
 
\subsection{Correlation of surface dose rates and pressure}

\begin{figure}
\centering
\noindent\includegraphics[width=0.80\columnwidth]{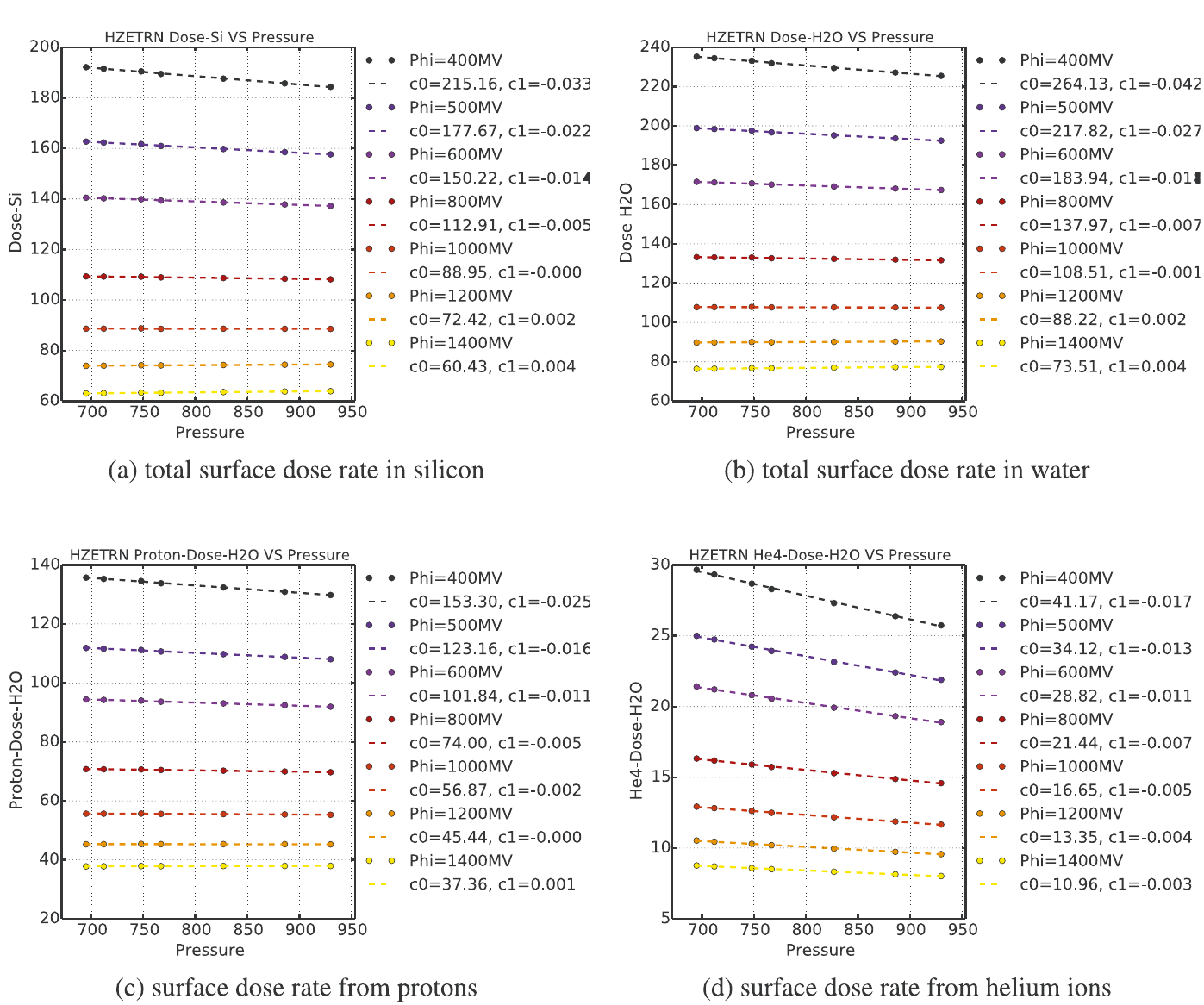}
\caption{HZETRN2015 results of the surface dose rates (y-axes, \textmu Gy/day) versus surface pressures (x-axes, Pascal). (a): total surface dose rate in silicon; (b): total surface dose rate in water; (c): surface dose rate from protons; (d): surface dose rate from helium ions. Results from different solar modulation potentials $\Phi$ (from top to bottom: 400, 500, 600, 800, 1000, 1200, 1400 MV) have also been plotted in different colors.  
The anti-correlation of the dose rate dependence on the pressure has been well fitted (dashed lines) for each $\Phi$ value with a linear function $y = c_0 + c_1 x$ where $c_0$ and $c_1$ are shown. 
 }\label{fig:HZETRN_doseVSpre_surface}
\end{figure}

In the calculations, we first place the detector (both silicon and water) on the surface of Mars under different surface pressures, $P$ [Pa], and different solar modulation potentials, $\Phi$ [MV]. 
The surface dose rates accumulated over all angles induced by protons, helium ions and all particles in water are shown in  Fig. \ref{fig:HZETRN_doseVSpre_surface} (c), (d) and (b) panels respectively. 
The result for dose rates recorded in a silicon detector is shown in Fig. \ref{fig:HZETRN_doseVSpre_surface} (a) and these values are generally lower than those detected by the water detector due to the smaller ionization energy in silicon.
The anti-correlation between the surface dose rate and the surface pressure as found by the RAD measurements \citep{rafkin2014} is present in the modeled data for small $\Phi$ values and is well fitted by a linear function as shown by dashed lines in the figure. 
 
This trend is also stronger for heavier particles due to their shorter mean free path and thus higher probability of getting fragmented when passing through the atmosphere. A comparison of proton induced dose rates to helium ions induced dose rates versus surface pressure is shown in Fig. \ref{fig:HZETRN_doseVSpre_surface} (c) and (d) respectively. 

Different colors in each panel stand for different solar modulation potentials from solar minimum to maximum conditions.
At solar minimum with smaller $\Phi$ values, when surface pressure and the total column depth increases, the dose rate decreases meaning that the Martian atmosphere, albeit very thin, is acting as a shielding layer against the incoming GCR doses.
However as the solar modulation potential $\Phi$ increases, the shielding effect becomes weaker and even vanishes for $\Phi \ge$ 1000 MV.
This is because solar modulation reduces primarily the lower-energy primary GCR flux, but hardly affects the high-energy GCRs which easily penetrate the Martian atmosphere as shown in Fig. \ref{fig:modulation}.

\subsection{Dose rates at different atmospheric depth}

The atmospheric column mass per area integrated from the top of the atmosphere (in units of $g/cm^2$) is an exact measure of pressure in a hydrostatic atmosphere with constant acceleration g which is 3.72 m/s$^2$ \citep{rafkin2014}. 
The surface pressure of e.g., 840 Pa thus can be transfered into column depth as 22.6 g/cm$^2$. 
The measurement of dose rates on the surface between pressure 700 Pa and 950 Pa is approximately equivalent to surface column depth of 18.9 and 25.7 g/cm$^2$ if only downward fluxes are considered. 
However, whether the linear anti-correlation of the surface dose rate measured by MSL/RAD versus surface pressure can be extrapolated to the upper altitudes of the atmosphere is still to be answered \citep{guo2015modeling}. 
In order to understand how the surface dose-pressure anti-correlation (at small solar modulation) differs from the vertical dose-depth correlation when the atmospheric altitude is much higher, we set the detectors in the model at different heights above the surface.  
We placed a stack of detectors at 4, 8, 12, 16, 17, 18 g/cm$^2$ elevation above the surface and thus for each pressure setup case of the Martian atmosphere, there are 7 different atmospheric depths. 
The resulting dose rate versus atmospheric depth from 7 different setups of the surface pressures, with $\Phi$ at 500 MV, is shown in Fig. \ref{fig:HZETRN_doseVSatmthk} with (a) for proton induced dose rate in water, (b) for helium ion induced dose rate in water, {(c) for dose rate induced by other types of particles,} (d) for total dose rate in water and (e) for dose equivalent rate [\textmu Sv/day].

It is shown that the proton induced dose rate and the total dose rate have a maximum peak at higher altitude of the atmosphere\footnote{This is not to be confused with the Pfotzer maximum which refers to the altitude of the maximum total flux, not the total dose rate, measured through different altitudes of the atmosphere. The column depth of the Pfotzer maximum on Earth has been measured to be at an altitude of about 20 km \citep{pfotzer1936} which corresponds to $\sim$ 100 g/cm$^2$. Measurements also suggest that this value is not constant and it depends on, e.g., the geomagnetic cutoff rigidity  \citep{moller2013}.}. 
This peak depth for proton dose rate is about 12 g/cm$^2$ due to the generation of secondary protons from e.g. higher energy protons and fragmentation  of heavier ions in the atmosphere. 
The peak depth for total dose rates is about 6 g/cm$^2$.
These results can be compared to previous stratospheric balloon measurement near the polar regions of Earth where the vertical cutoff rigidity is about 0.5 GV \citep{moller2013}. Although their measurement didn't have any data for atmopheric depth lower than 20 g/cm$^2$, the dose rate showed a flattening trend towards the lower atmospheric depth. 
Therefore the anti-correlation between dose rates and pressure measured near the surface, also shown in Fig. \ref{fig:HZETRN_doseVSpre_surface}, should not be extrapolated to the top of the atmosphere. 
{We have also plotted the dose rate induced by particles other than protons and helium ions (including heavier primary GCR ions and other secondaries) as shown in Fig. \ref{fig:HZETRN_doseVSatmthk}(c) which seems to show a nadir around the depth of the dose peak.
At altitudes above this depth ($\le \sim 6$ g/cm$^2$), the high-charge primary GCR ions contribute mainly to the dose rate in (c). As the atmospheric depth grows, these primary particles are shielded and their flux and dose rate decreases as shown in (c). }
Meantime, secondaries are being generated from high energy ions to lower-energy lower-charged particles. 
The contributed dose rate by the primaries decreases and by the secondaries increases. 
The net effect of these opposing changes is an increase in the {total} accumulated dose rate shown in (d). 
{At altitudes deeper than $\sim 10$ g/cm$^2$, primary GCRs which have fragmented in the atmosphere contribute much less to the dose rate in (c) while atmospheric and albedo secondaries start playing a more important role. 
Therefore as atmospheric depth increases, the dose rate in (c) slightly increases. 
However, the total dose rate shown in (d) is still largely dominated by proton and helium ion induced dose rate which experiences the atmospheric shielding effect and anti-correlates with the depth.}

The dose rate induced by helium ions shows a decaying curve versus the atmospheric column depth indicating that the shielding of the helium ions dominates over the generation of secondaries, due to their larger charges and smaller mean free path. 
As shown in Fig. \ref{fig:HZETRN_doseVSatmthk}(b), the data have been fitted by the following decay exponential function for seven different cases of total surface pressures:
\begin{eqnarray}\label{eq:flux_attenuation}
D_{He}(\sigma) = D_{0_{He}} \exp(- \frac{\sigma}{\lambda_{He}}), 
\end{eqnarray}
where $D_{0_{He}}$ is the helium dose rate peak which is close to the top of the atmosphere, $\sigma$, in units of $g/cm^2$, is the atmospheric column mass per area integrated from the altitude of $D_{0_{He}}$ and $\lambda_{He}$ is its characteristic shielding depth and the overall fitting through all the data points results in $\lambda_{He}$ of about 41.2 g/cm$^2$.
It is interesting to notice that as the surface pressure increases, the exponential fitting shows a slight increase of $\lambda_{He}$ meaning a weaker shielding effect, likely due to the increased contribution of secondary helium ions in the atmosphere. 
The exponential fit is merely an empirical approximation where the shielding effect is much stronger than the generation of secondaries and should be applied with caution as input parameters (primary particle energies and types) and setup conditions (atmospheric depth and directions for integration) vary.

The dose equivalent rate doesn't show a peak as the dose rate does at high altitudes and it declines continuously and nonlinearly as the column depth increases. 
This is more related to the fact that the dose equivalent rate, compared to the dose rate, has a higher contribution from heavy ions which fragment more as they go through the atmosphere {as also shown in Fig. \ref{fig:HZETRN_doseVSatmthk}(c) at small column depth} and thus the shielding effect dominates even more. 
However as the total dose equivalent rate is a combination of all particles (both primaries and secondaries) in all directions, a shielding-driven exponential function doesn't fit the data very well. 

\subsection{The effect of solar modulation on dose-depth correlations}

The dependence of the above depth effect on dose rates has also been tested under different solar modulation potentials as shown in Fig. \ref{fig:HZETRN_doseVSatmthk_phi}. 
As the solar modulation becomes stronger (7 different $\Phi$ values from top to bottom lines are 400, 500, 600, 800, 1000, 1200, 1400 MV in each panel), the primary GCR flux decreases especially at lower energy ranges as shown in Fig. \ref{fig:modulation}. 
Therefore the proton dose rate reaches its peak at a deeper depth due to the delay of accumulating of low-energy secondary protons in the atmosphere from high-energy protons and heavy ions. 
For $\Phi$ at bigger values above 800 MV, it seems that the proton dose peak is on/below the surface of Mars. 

Similar influences of solar modulation are also shown in the total dose rate in water in Fig. \ref{fig:HZETRN_doseVSatmthk_phi}(c). 
For conditions with small $\Phi$ values when there is a great amount of low-energy primaries, the dose rate decreases as the atmospheric depth increases due to the shielding of such primaries over-weighing the generation of secondaries by high energy GCRs. 
For bigger $\Phi$ values, the low-energy ends of the GCRs are greatly reduced and the secondary fragmentation (which increases with depth) from high energy particles weighs more than primaries in terms of dose contribution. E.g., for $\Phi$ at 1400 MV shown as the bottom line, dose rate increases as the atmospheric depth increases indicating that the dose peak may be on/below the surface of Mars.

For helium ion induced dose rates shown in Fig. \ref{fig:HZETRN_doseVSatmthk_phi}(b), the dose-depth dependence becomes weaker for larger solar modulation potentials due to the reduce of lower-energy particles in the primary flux which are more responsible for the depth effect. 
This trend of solar modulation effects is also visible for the dose equivalent rate, shown in Fig. \ref{fig:HZETRN_doseVSatmthk_phi}(c), which differs from the dose rate by larger contributions of heavy ions. 
    
\section{Martian surface measurements by MSL/RAD}\label{sec:rad_obs}
Since the successful landing of the Curiosity rover in Gale crater in August, 2012, the MSL/RAD instrument has been conducting the first-ever in-situ measurements of the Martian surface radiation \citep{hassler2014}.
The surface pressure which is a direct measurement of column density is also recorded on board by the Rover Environmental Monitoring Station (REMS) \citep{haberle2014preliminary}. 
The heliospheric solar {activity} has been measured for decades via e.g. neutron monitors at Earth and {its modulation of the GCR flux can be parameterized as modulation potential $\Phi$ whose values are often derived for each Carrington rotation representing the averaged heliospheric condition} \citep[e.g.,][]{usoskin2005}. 
About four years (August 2012 -- August 2016, $\sim$ 2 Martian years) of surface radiation data, surface pressure, and solar modulation $\Phi$ have been employed herein for our modelling purpose. 

\subsection{Measured dose rates, surface pressure and heliospheric $\Phi$}

The radiation doses on the surface of Mars {from all directions}, contributed by both primaries and secondaries, both charged and neutral particles, are measured simultaneously in two detectors of RAD: the silicon detector B and the plastic scintillator E.
Viewed from top to bottom, the RAD sensor head consists of a  stack of silicon detectors, namely, A,B,and C followed by a Tl-doped CsI scintillator crystal (D) and a tissue-equivalent plastic scintillator (E). Both D and E are enclosed in an efficient plastic scintillator anti coincidence (F1 on the side and F2 at the bottom). 
Detector E has a composition similar to that of human tissue and water and measures a higher dose rate value than detector B due to the difference of ionization potential in silicon and plastic. 
Because of the bigger size of the E detector, the dose rate it measures shows much better statistics than the dose rate measured in B.
For more details of the RAD instrument design, please refer to \citet{hassler2012}.

The surface pressure at Gale crater is recorded at high time resolution by MSL/REMS, and it evolves regularly and concurrently at both diurnal and seasonal time scales \citep{rafkin2014, guo2015modeling}. 
The diurnal variation of pressure is caused by the thermal tide at Gale Crater and the day and night column mass oscillates about $\pm 5\%$ relative to the median \citep{haberle2014preliminary}.
The seasonal atmospheric pressure variation is controlled by a complex balance between the cold and warm poles \citep[e.g.,][]{tillman1988mars} and in Gale Crater it varies by 25 \% over the course of one Martian year.

On the other hand, solar modulation potential $\Phi$ is an approximate index of heliospheric modulation that generally varies slowly over the course of the solar cycle, but can also undergo rapid changes due to fast-varying solar activity including solar particle events. 
Therefore $\Phi$ is often treated as an average over one Carrington rotation (about 26 sols at Mars orbit). 
$\Phi$ can be derived at Earth using e.g., Oulu neutron monitor count rate data \citep[e.g.,][]{usoskin2005}; Precise $\Phi$ measurements at Mars, however, are not available.  
Assuming that the modulation condition in the heliosphere during each Carrington rotation is stable and uniform across the different heliospheric longitudes, we can approximately evaluate the average $\Phi$ for each rotation period at Earth and extrapolate it to Mars orbit considering the radial gradient of the modulation from 1AU to 1.5 AU \citep{schwadron2010earth, guo2015modeling}. 

\subsection{Fitting dose rate and pressure correlation}

We use the method described in \citet{rafkin2014} to produce the average diurnal perturbations of the dose rate resulting from the pressure changes; this approach aims at isolating the diurnal pressure-responsible variations in the RAD measurements from other disturbances of daily variations such as solar particle events and Forbush decreases. 
The anti-correlation of the dose rate perturbations versus pressure changes is shown in Fig. \ref{fig:hourlyPert_doseE} where the x-axis shows the mean perturbation of pressure in each hour and y-axis shows the mean perturbation of dose rate in each hour. 
The mean perturbation of data in each hour represents the mean of the difference between hourly dose rate and its {corresponding} daily mean.
The mean perturbation of dose rate, $\bar{\delta D_{h}}$, can be readily correlated with the hourly pressure perturbation, $\bar{\delta P_{h}}$, and their relationship follows a clear anti-correlation which can be fitted with a first-order polynomial function:
\begin{eqnarray}\label{eq:hourly_pert_correlation}
\bar{\delta D_{h}} = \kappa \cdot \bar{\delta P_{h}}.
\end{eqnarray}
The {resulting} parameter $\kappa$ [\textmu Gy/day/Pa] and the linear fits are also shown in each panel of Fig. \ref{fig:hourlyPert_doseE}.

The top and bottom panels show the results from plastic E and silicon B measurements respectively. The error bars for the plastic measurements are much smaller due to the larger geometric factor of the plastic detector. 
The data are taken during two different periods when the solar modulation potentials were very different. The left panels show data taken from 2013-5-23 to 2014-4-14 when the averaged $\Phi$ at Earth measured by Oulu Neutron monitor is about 634 MV. Accounting for the radial distance from 1AU to 1.5 AU (Mars' orbit), we correct this value to be about 578 MV. 
$\kappa$ fitted during this period is about $-0.13 \pm 0.02$ and $-0.12 \pm 0.07$ for plastic and silicon detectors respectively.
The right panels contain data taken from 2015-5-30 to 2016-8-3 when the averaged $\Phi$ at Earth is about 537 MV and 489 MV at Mars' orbit.
$\kappa$ fitted during this weaker solar modulation period is about $-0.17 \pm 0.03$ and $-0.17 \pm 0.08$ for plastic and silicon detectors respectively.
It is readily shown that anti-correlation coefficient $\kappa$  has a larger absolute value under weaker solar activities, agreeing fairly well with the modeled results shown in Fig. \ref{fig:HZETRN_doseVSpre_surface}.

The absolute values of ${\kappa}$ obtained from models are in the range of [0.014, 0.033] for silicon detector and [0.018, 0.042] for water detector with $\Phi$ changes from 600 to 400 MV. 
These values are, however, substantially smaller compared to the results from measurements. 
This might be due to the following reasons related to the details of how the instrument detects particles and measures dose rates:
\begin{itemize}
\item{The dose rate obtained from the measurements are from particles which make it through the shielding of the rover, the electronic box and eventually to the plastic/silicon detectors.  
E.g., for a downward proton to pass through the thick detector D and reach the plastic detector E, it has to have an energy above $\sim$ 100 MeV; And relativistic primary particles with higher energies would lose energy before it reaches E, leading to a very different energy deposition pattern in E compared to the same particles that would reach the detector unhindered. 
The shielding around the detectors, however, is highly non-uniform and such modulation of the original surface spectrum differs at different incident angle of the particles.
In general, we estimate the shielding filters out/dilutes a good amount of low energy secondaries produced in the atmosphere which respond positively, rather than negatively, to the pressure changes. These particles are all included in the calculations of HZETRN2015 and have resulted in a smaller anti-correlation coefficient.}
\item{In the modeling process, the particle spectra were converted to corresponding dose rates via an analytic function describing the ionization energy deposit of particles in certain materials, e.g., the Bethe-Bloch formula which is a function of the linear energy transfer dE/dx versus incoming particle energy E. 
In reality, however, the incoming particles, especially those with higher energies and bigger charges, have a probability to interact with the detectors and produce low-energy secondaries that may deposit more energy in the detector than the original particle and, thus, contribute more to the measured dose. 
This consequently enhances the dose rate contribution by heavy ions which are much more responding to the shielding of the atmosphere and thus results in our bigger estimations of the $\kappa$.}
\end{itemize}

In fact, a recent attempt of comparing RAD measured particle spectra and different model predictions of GCR spectra on the surface of Mars has shown quite some discrepancies between modeled and measured results \citep{matthia2016martian}. This could be partly due to similar reasons listed here and the local shielding environment around the RAD detector has modified the original surface spectra making the direct comparison very difficult. 
Further calculations and/or simulations accounting for the shielding of the outer detectors, electronic box and even the complex rover body as well as the production of secondaries inside the detectors could be carried out for better quantitative determination of the above factors. 

\subsection{The variation of the dose rate and pressure correlation}

It is already shown in Fig. \ref{fig:hourlyPert_doseE} that different $\Phi$ may result in different dose rate-pressure correlation coefficients. 
To analyze this solar modulation effect quantitatively, we employ the data collected over nearly four years of mission period (from August 2012) as shown in Fig. \ref{fig:doseE_pre_phi}.  

The solar modulation potential at Mars distance extrapolated from Oulu Neutron Monitor measurement \citep{usoskin2005} is plotted in green (right y-axis) and binned into 26 sols which is the Carrington rotation period at Mars. 
The error bars stand for the standard deviations of data within the each period.
The modulation potential changes irregularly with big uncertainties and has a range from 400 to 700 MV over the 4 year period. 

The dose rate measured in plastic detector E is shown in black (left y-axis) and the surface pressure data recorded by REMS is shown in red (right y-axis). 
Their anti-correlation factor $|\kappa|$ fitted by Eq. \ref{eq:hourly_pert_correlation} for each 26 sols is shown in blue with its values scaled up by 1000 times for better visualization.
The plastic dose rate measured on the surface of Mars integrated over all directions ranges between about 190 and 260 \textmu Gy/day within the time period of the measurement.
The Martian seasonal cycle during the 2 Martian years is visible in the pressure data. The error bar of the pressure data also includes the actual diurnal-oscillation due to the surface thermal tide. 
It is obvious that the dose rate and solar modulation are anti-correlated, but the seasonal pressure influence must also be taken into account when analyzing the long-term variation of the dose rate \citep{guo2015modeling}.

The anti-correlation between pressure and dose rate, $|\kappa|$, fitted by Eq. \ref{eq:flux_attenuation}, also shows a negative correlation with $\Phi$, i.e., larger $|\kappa|$ during weaker solar activities. 
The linear regression correlation coefficient between $|\kappa|$ and $\Phi$ is -0.67 which is significant but not a very strong correlation due to the big uncertainties in the data. 
$|\kappa|$ and $\Phi$ can be fitted by a linear function $-\kappa = c_0  + c_1 \cdot \Phi$ which is also shown as a red line in the right panel of Fig. \ref{fig:doseE_pre_phi_fit}. 
Other functions may also be employed for the fitting but we don't find another function describing the data better and/or being more physical within the limited range of the parameter range. 
The fitted parameters for the linear function is $c_0 = 0.28 \pm 0.03$ \textmu Gy/day/Pa and $c_1 = -2.9 \pm 0.6 \times 10^{-4}$ \textmu Gy/day/Pa/MV. 
For typical values of the current solar modulation $\Phi$ at 400, 500, and 600 MV, the above function results in $-\kappa$ about 0.164, 0.135 and 0.106 \textmu Gy/day/Pa. 
At $\Phi$ about 965.5 MV, $\kappa$ decreased to zero meaning the anti-correlation between surface dose rate and pressure changes vanishes for stronger solar activities than this value, also agreeing well with the modeling results, shown in Fig. \ref{fig:HZETRN_doseVSpre_surface}, where the shielding effect disappears at about 900-1000 MV. 
The same analysis for $\kappa$ versus $\Phi$ correlation has been applied to the modeled results for the case of total water dose rate in the range of $\Phi$ from 400 to 700 MV. 
The resulted fitting parameters are $c_0 = 0.072$ and $c_1 = -8.6 \times 10^{-5}$. 
These values are smaller than the ones obtained from measurements due to the same reasons addressed in the last section. 
Although the quantitative comparison of $\kappa$ versus $\Phi$ correlation between measurements and models is not satisfactory, the qualitative results are both sensible and agree with each other fairly well.   

It is also visible in Fig.\ref{fig:doseE_pre_phi_fit} that $\kappa$ is slightly anti-correlated with pressure. This indicates that as pressure increases the attenuation effect slightly decreases, similar to the behavior of the helium ion particle dose rates shown in Fig. \ref{fig:HZETRN_doseVSpre_surface}, likely caused by the increased contribution of secondaries in the atmosphere. 
Due to large uncertainties of the data, an over-simplified linear fit was not carried out for $\kappa$ -- pressure correlation avoiding over-interpretation of the results.

\section{Discussions and conclusions}
In order to study the concurrent influences of radiation dose rate by both the solar modulation and Martian diurnal as well as seasonal pressure variations, it is important to separate the pressure-driven perturbations from the solar modulation in the variation of the GCR-induced dose rates on the surface of Mars. 
\citet{guo2015modeling} assumed, however, independent pressure and solar modulation effects on dose rate and analyzed the quantitative anti-correlation between dose rate and pressure or $\Phi$. 
The empirical fitting of dose-depth correlation therein is valid for a small range of $\Phi$ variations as the analysis was limited by the data obtained by then. 

In the current study, we use the most up-to-date dose rate and pressure data collected in the past 4-year mission period of MSL on the surface of Mars. The solar modulation has also become much weaker in the past year allowing a much wider range of parameter studies hereby.  

Moreover, we employed extensive HZETRN2015 calculations to investigate the atmospheric effect on dose rate at the surface of Mars as well as at higher altitudes above the surface. 
The input GCR spectra were obtained based on the Badhwar-O'Neill 2010 model with particle charges ranging up to 28.
These particle spectra are then used as inputs for the HZETRN2015 model to generate the particle spectra for different ion species which are then converted into accumulated dose rates. 
Seven virtual detectors are located in the model for recording the integrated dose rates: one at the surface and the others at elevations of 4, 8, 12, 16, 17 and 18 g/cm$^2$ above the surface. 
Seven different surface pressures, in a range of typical Martian surface pressures measured by MSL/REMS at Gale Crater through different seasons, are considered in the model. 

The modeled results are then compared with MSL/RAD dose rate data on the surface of Mars during the past four years of measurements. 
A summary of the main results from the calculations and the measurements is as following: 
\begin{itemize}
\item{The GCR-induced surface dose rate variation is driven by both the solar modulation and Martian atmospheric pressure changes.}
\item{In the long-term, the solar modulation has a much stronger effect on the dose rate variations.}
\item{The surface dose rate is anti-correlated with the surface pressure (atmospheric depth) for solar modulation potentials smaller than $\sim$ 900-1000 MV, as shown by the model and indicated by the measurement.} 
\item{As suggested by modeled results, this dose-depth anti-correlation (under small $\Phi$ values) could be extrapolated close to the altitude of the dose peak but not to the top of the atmosphere.}
\item{The dose peak shown in the model varies as $\Phi$ changes. 
It appears at deeper atmosphere (close to the surface) under stronger solar activities and vice versa.}
\item{As solar modulation varies, the dose-depth anti-correlation also changes. At smaller $\Phi$ values, this anti-correlation is stronger and vice versa. This is due to the presence of more lower-energy {GCR primary} particles, at weaker solar activities, which are more affected by the atmospheric shielding.}
\item{The dose-pressure correlation factor $|\kappa|$ obtained from HZETRN2015 ranges between [0.018, 0.042] \textmu Gy/day/Pa for a water detector with $\Phi$ varying from 600 to 400 MV. 
However, $|\kappa|$ derived from measured data from the plastic detector, for similar $\Phi$ values, is in the range of [0.10, 0.20] \textmu Gy/day/Pa, much larger than that from models likely because the measurement may have a reduced contribution from low-energy secondaries in the atmosphere and a relatively enhanced contribution induced by heavier ions producing secondaries in the detectors.}
\item {$\kappa$ can be anti-correlated with $\Phi$ since the shielding effect decreases as solar modulation becomes stronger. The linear fit of $\kappa$ versus $\Phi$ suggests that the shielding effect may vanish as $\kappa$ approaches zero at large $\Phi$ values $\sim$ 900-1000 MV.  }
\end{itemize}  

In summary, the current paper analyzed atmospheric depth effect on the variations of the radiation dose rate and how this effect changes as solar modulation varies or surface pressure differs. 
Modeling results indicate that the atmospheric shielding effect which MSL/RAD has seen in the past four years may be due to the weak/medium solar modulation during this period. 

According to recent solar cycle models \citep[e.g.,][]{kapyla2016multiple}, we may be at the start of a grand solar minimum and the solar modulation in future years could be even weaker than the current measurements. 
Therefore for future human exploration to planet Mars during solar minimum periods, it is important to {take into consideration} the atmospheric shielding effect. 
Based on the RAD measurements, a first-order estimation of $\kappa$ at $\Phi = 200$ MV would be 0.222 \textmu Gy/day/Pa which could result in about 55.5 \textmu Gy/day of dose rate difference between minimum and maximum seasonal pressure conditions ($\sim$ 700 Pa and 950 Pa) at Gale crater. 
This is about 25\% of the average dose rate ($\sim$ 220 \textmu Gy/day) measured so far. 
This suggests that it would be better to avoid the minimum pressure season of the southern hemisphere late winter caused by the southern CO$_2$ ice cap reaching its maximal extent \citep[e.g.,][]{tillman1988mars}.
 
\begin{figure}
\centering
\noindent\includegraphics[width=0.80\columnwidth]{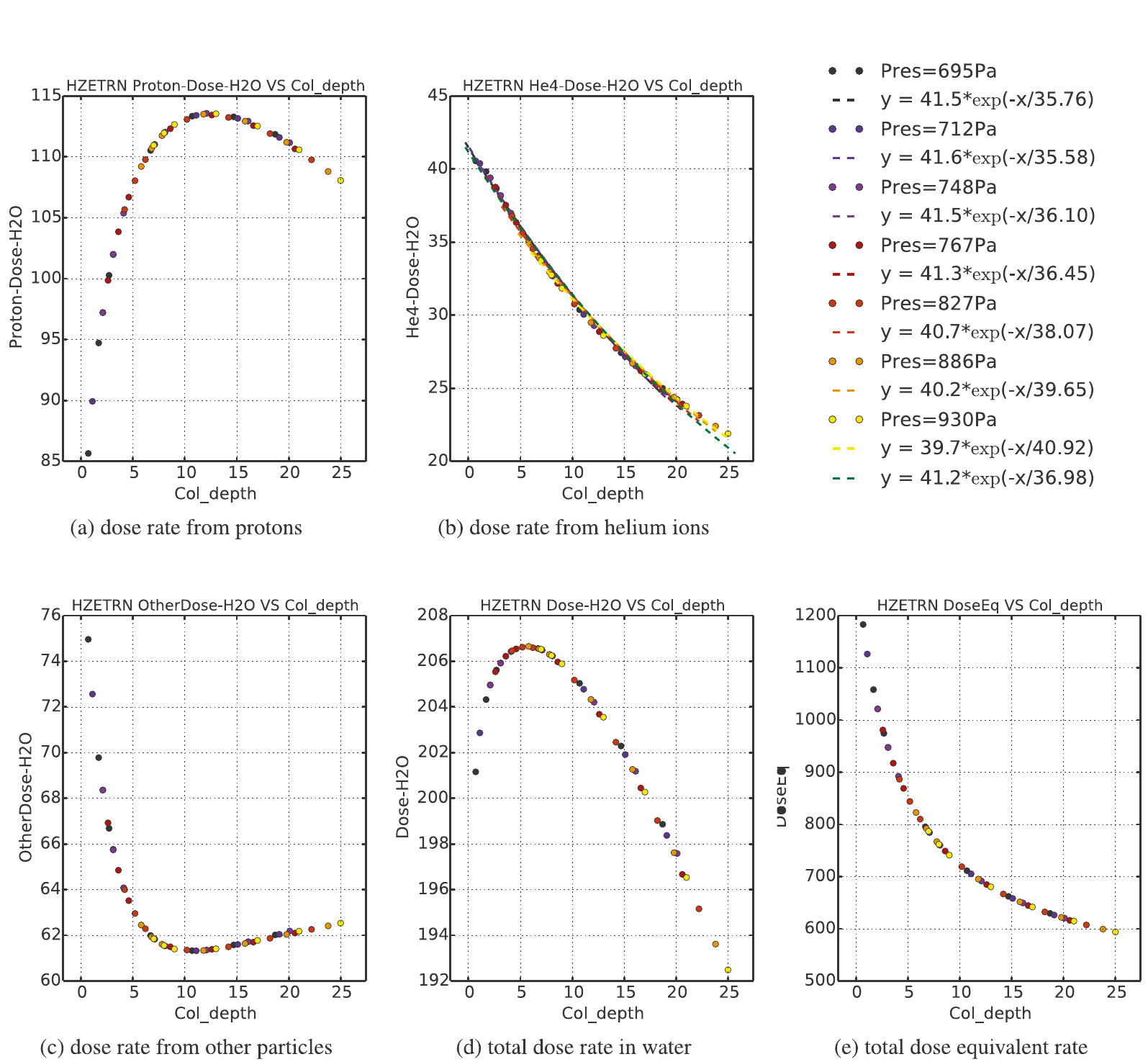}
\caption{HZETRN2015 results of dose rates (\textmu Gy/day) in water from protons (a), helium ions (b), particles other than protons or $^4$He (c), as well as total dose rate (d) and total dose equivalent rate (\textmu Sv/day, e) versus atmospheric depths (x-axes, g/cm$^2$) at $\Phi = 500$ MV. 
The atmospheric depths are also from different setups of surface pressures which are indicated by different colors. 
The exponential fit in (b) for each different surface pressure setup is also shown on the right side of the panel. An overall fit of all the data is shown in green. }\label{fig:HZETRN_doseVSatmthk}
\end{figure}

\begin{figure}
\centering
\noindent\includegraphics[width=0.8\columnwidth]{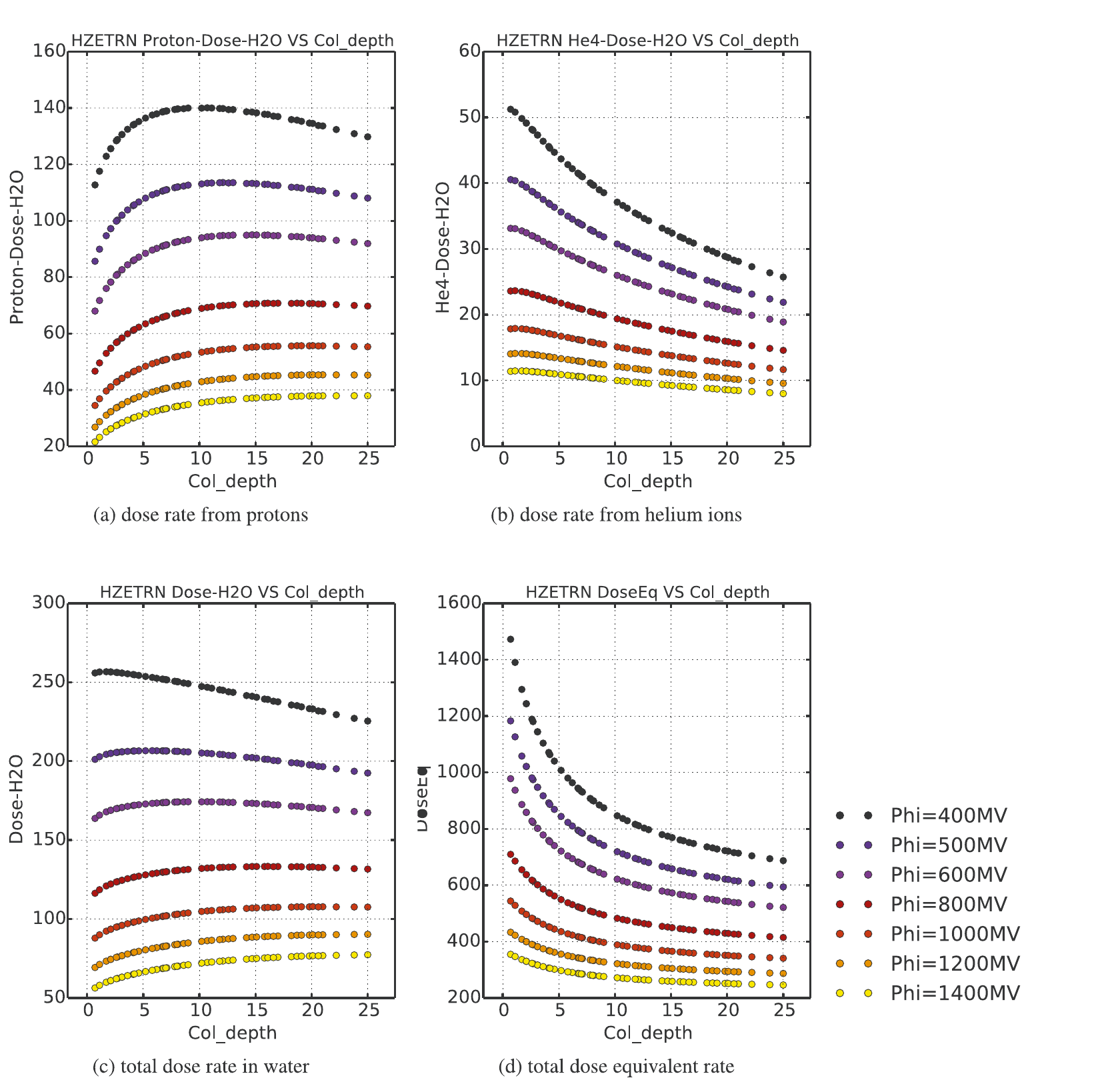}
\caption{HZETRN2015 results of dose rates (\textmu Gy/day) from protons (a) and helium ions (b) in water as well as total dose rate (c) and total dose equivalent rate (\textmu Sv/day, d) versus atmospheric column depths (x-axes, g/cm$^2$) for different $\Phi$ values shown in different colors. 
}\label{fig:HZETRN_doseVSatmthk_phi}
\end{figure}

\begin{figure}
\centering
\noindent\includegraphics[width=0.8\columnwidth]{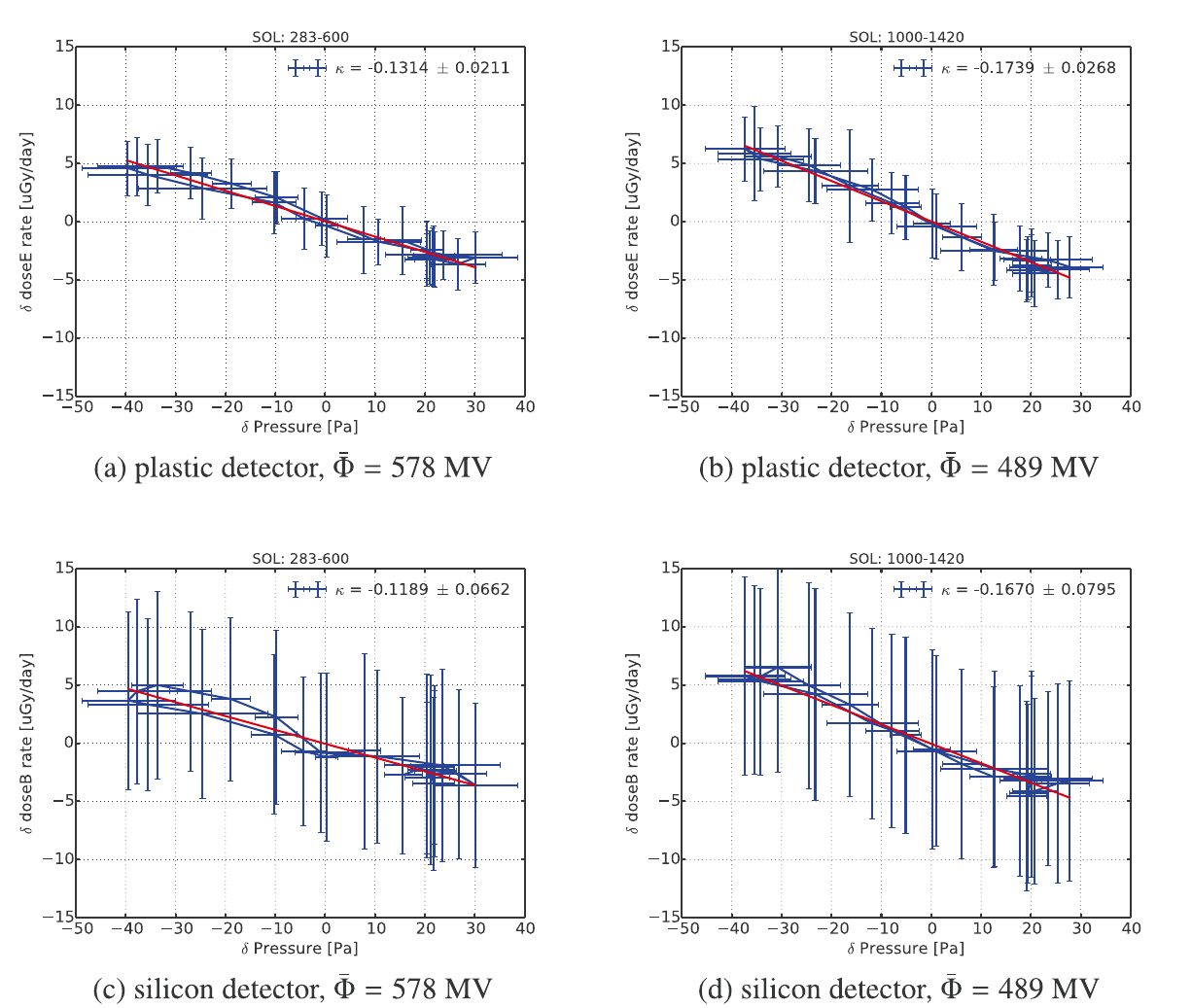}
\caption{Hourly perturbations of dose rate $\bar{\delta D_{h}}$, measured by different detectors(top: plastic detector; bottom: silicon detector), versus pressure $\bar{\delta P_{h}}$ under different solar modulation potential $\Phi$ (left: $\bar{\Phi}=578$ MV and right  $\bar{\Phi}=489$ MV). 
The error bars stand for the standard deviation of the averaged hourly perturbation.
The fitted anti-correlation is shown as a red line with a slope of $\kappa_d$ in units of \textmu Gy/day/Pa.} 
\label{fig:hourlyPert_doseE}
\end{figure}

\begin{figure}
\centering
\includegraphics[trim=10 10 10 10,clip, width=0.70\columnwidth]{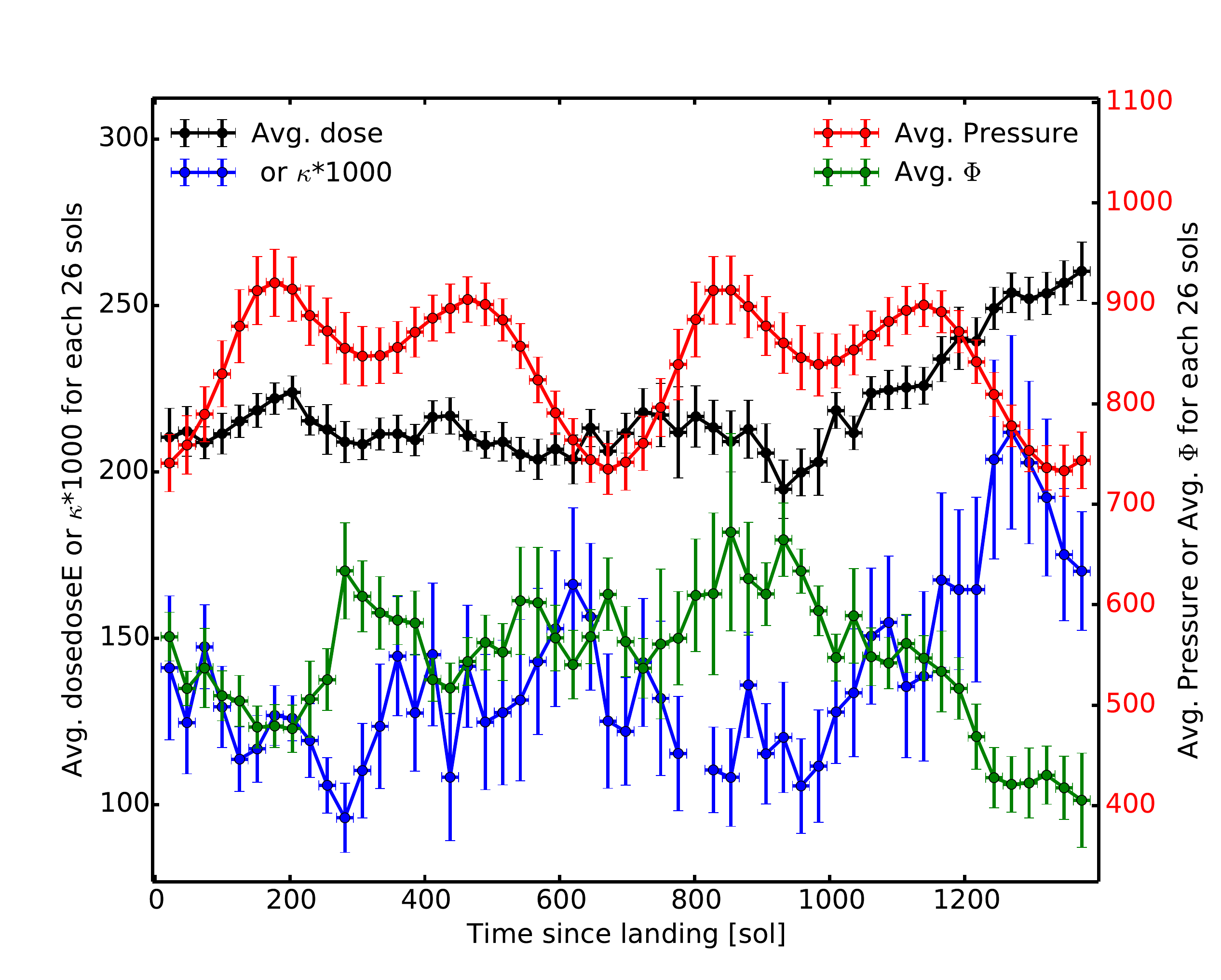}
\caption{Dose rate, pressure, and $\Phi$ data collected in the time range from 14 August 2012 until 29 June 2016. Left y-axis: 26-sol binned RAD plastic dose rate (\textmu Gy/day, black) and derived pressure-dose rate correlation $|\kappa|$ (in unit of 1000*\textmu Gy/day/Pa, blu;  Right y-axis: 26-sol binned surface pressure (Pascal, red) and $\Phi$ (MV, green) at Mars' orbit.}\label{fig:doseE_pre_phi}
\end{figure}

\begin{figure}
\centering
\includegraphics[trim=100 0 8 40,clip,  width=0.70\columnwidth]{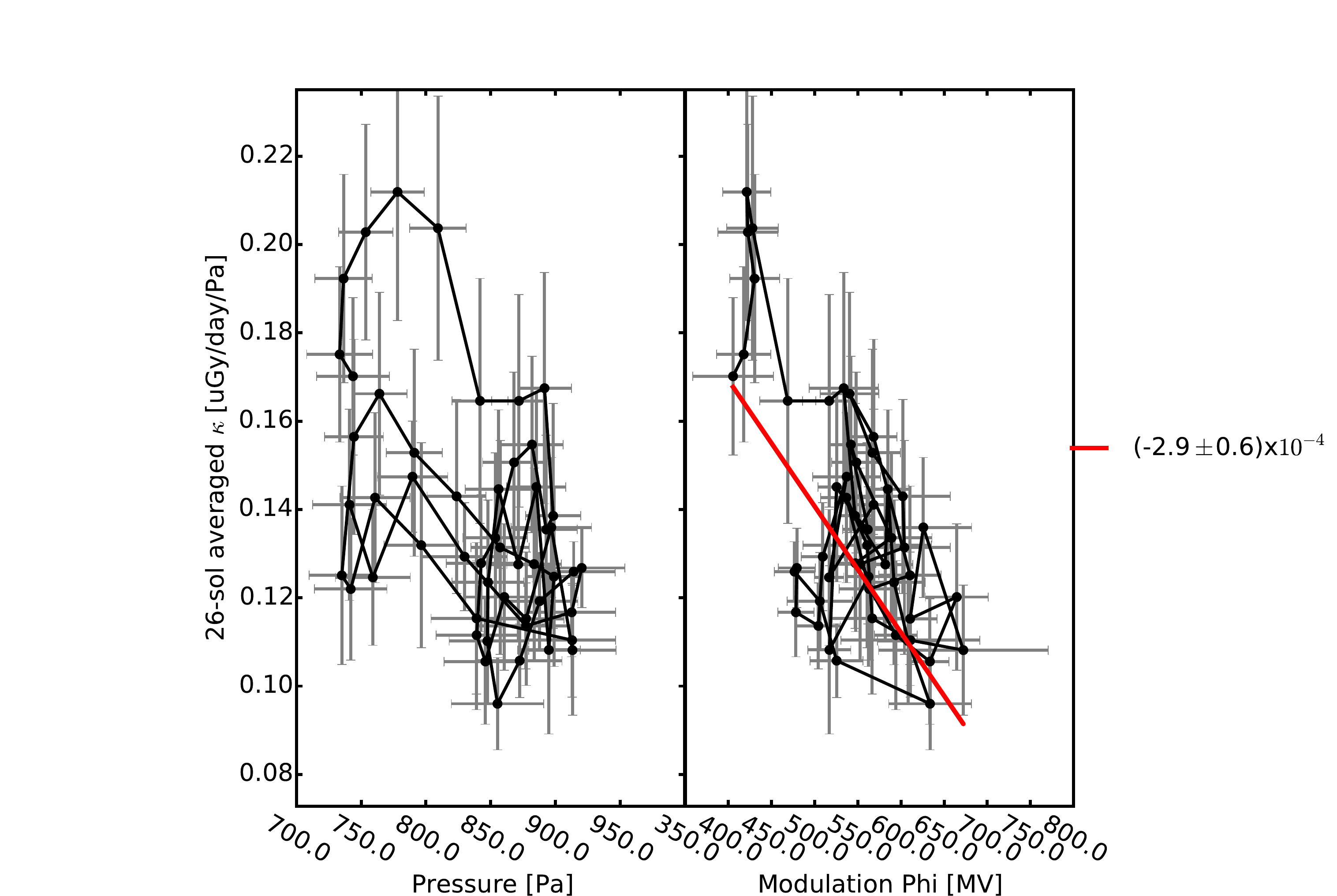}
\caption{Derived pressure-dose rate correlation $\kappa$ (y-axis, \textmu Gy/day/Pa) for each 26 sols in the time range from 14 August 2012 until 29 June 2016 versus pressure (left panel, Pascal) and $\Phi$ (right panel, MV). In the right panel $|\kappa|$ and $\Phi$ has been fitted by a linear function (red line) and the coefficient is 2.9 $\pm$ 0.6 $\times$ 10$^{-4}$ \textmu Gy/day/Pa/MV. }\label{fig:doseE_pre_phi_fit}
\end{figure}

In terms of biological effectiveness, the dose equivalent rate is often more referred to for {evaluating} the deep space exploration risks \citep{sievert1959}. 
In fact, the relative difference of dose equivalent rate between different seasons would be even bigger since heavier ions whose fluxes are more affected by the atmosphere have a bigger contribution to dose equivalent than to dose rate. 
From measurement, dose equivalent rate is estimated by multiplying the dose rate by an average quality factor $<Q>$ which is determined through the LET histogram of the measured particles \citep{icrp60}. 
The estimated $<Q>$ on the surface of Mars for the first 300 days of measurement was about 3.05 $\pm$ 0.3 \citep{hassler2014}, considerably smaller than 3.82 which was measured during the cruise phase \citep{zeitlin2013} where there was less shielding by the spacecraft on average. 
Based on dose equivalent rate and dose rate values from the HZETRN2015 model as shown in Figs. \ref{fig:HZETRN_doseVSatmthk} and \ref{fig:HZETRN_doseVSatmthk_phi}, we can also derive $<Q>$ from the model\footnote{Note that HZETRN2015 model calculates the dose equivalent directly from primary particle types and energy spectra without using $<Q>$. We here estimate $<Q>$ for the purpose of comparing with measurements.}.
At $\Phi = 500$ MV which is an approximation during periods studied in \citet{hassler2014}, $<Q>$ derived from the HZETRN2015 model is 3.08 and 3.17 for boundary pressure conditions and this is consistent with that from the measurement within uncertainties \citep{hassler2014}. 
The difference of dose rates of two pressure boundaries is 33.75 \textmu Gy/day which is about 15\% of the total average. 
Folding with $<Q>$, the resulting dose equivalent rate difference is $\approx$ 126 \textmu Sv/day or about 19\% of the total average. 
In our future work, we will try to derive $<Q>$ at different atmospheric and solar modulation conditions and thereby obtain the correlation of dose equivalent rate with column depth.  
   
At stronger solar modulation conditions, the atmospheric influence is however much weaker since the primary GCRs would have fewer particles responding to the atmospheric changes. 
At very large $\Phi$ values, a deeper atmosphere may even slightly enhance the total dose rates resulting in {a} positive correlation between dose rate and surface pressure (or column depth). 
Data to be collected at solar maximum conditions will be necessary to test the above hypothesis.   
Furthermore, the complex shielding around the dose detectors and how this affects our measurement of dose rate as well as its atmospheric response will be investigated in more detail using full Monte-Carlo simulations.

\begin{acknowledgments}
RAD is supported by the National Aeronautics and Space Administration (NASA, HEOMD) under Jet Propulsion Laboratory (JPL) subcontract \#1273039 to Southwest Research Institute and in Germany by DLR and DLR's Space Administration grant numbers 50QM0501 and 50QM1201 to the Christian Albrechts University, Kiel. 
Part of this research was carried out at JPL, California Institute of Technology, under a contract with NASA. 
We are grateful to the Cosmic Ray Station of the University of Oulu and Sodankyla Geophysical Observatory for sharing their Neutron Monitor count rate data. 
The data used in this paper are archived in the NASA Planetary Data System’s Planetary Plasma Interactions Node at the University of California, Los Angeles. The archival volume includes the full binary raw data files, detailed descriptions of the structures therein, and higher-level data products in
human-readable form. The PPI node is hosted at http://ppi.pds.nasa.gov/.
\end{acknowledgments}


\end{article}

\end{document}